\documentclass[11pt,a4paper]{article}
\usepackage[hyperref]{tto2019}
\usepackage{times}
\usepackage{balance}
\usepackage{latexsym}
\usepackage{moresize}
\usepackage{graphicx}
\usepackage{multirow}
\usepackage{url}

\aclfinalcopy 

\title{A Study of Cyber Hate on Twitter with Implications for Social Media Governance Strategies}
\author{Rob Procter\\
  Department of Computer Science\\
  Warwick University\\
  \small{\texttt{rob.procter@warwick.ac.uk}}
  \And
 Helena Webb Marina Jirotka\\
  Department of Computer Science\\
  Oxford University\\
  \small{\texttt{helena.webb$|$marina.jirotka@cs.ox.ac.uk}}
  \AND
Pete Burnap\\
 Department of Computer Science\\
 Cardiff University\\
 \small{\texttt{BurnapP@cardiff.ac.uk}} \\\And
William Housley 
Adam Edwards
Matt Williams\\
School of Social Sciences\\
  Cardiff University \\
  \small{\texttt{HousleyW$|$EdwardsA2$|$WilliamsM7@cardiff.ac.uk}}\\}

\date{}

\usepackage[]{hyperref}
\hypersetup{
pdftitle={SIGCHI Conference Proceedings Format},
pdfauthor={LaTeX},
pdfkeywords={SIGCHI, proceedings, archival format},
bookmarksnumbered,
pdfstartview={FitH},
colorlinks,
citecolor=black,
filecolor=black,
linkcolor=black,
urlcolor=black,
breaklinks=true,
}

\begin{document}

\maketitle

\begin{abstract}
This paper explores ways in which the harmful effects of cyber hate may be mitigated through mechanisms for enhancing the self-governance of new digital spaces. We report findings from a mixed methods study of responses to cyber hate posts, which aimed to: (i) understand how people interact in this context by undertaking qualitative interaction analysis and developing a statistical model to explain the volume of responses to cyber hate posted to Twitter, and (ii) explore use of machine learning techniques to assist in identifying cyber hate counter-speech. 
\end{abstract}


\section{Introduction}
In this paper we report on a study that was conducted as part of the `Digital Wildfires' project \cite{webb2015digital,webb2016digital} on the efficacy of 
self-governance in social media through the agency of counter-speech. 


`Cyber hate' (i.e., postings that are offensive, threatening and intended or are likely to stir up hatred against individuals or groups) \cite{awan2014islamophobia,burnap2016us,titley2014starting} can inflict severe harm on individuals, groups and communities. 
This has led to discussion –- in both academic \cite{webb2015digital} and policy fields \cite{titley2014starting,gagliardone2015countering} –- of whether it is (a) desirable and (b) feasible to enact governance mechanisms to regulate the conduct of social media users and so limit its impact. As a result, there has been a growing interest in examining whether self-governance through counter-speech -- ``a common, crowd-sourced response to extremism or hateful content'' \cite{bartlett2015counter} -- in online discussions can reduce the spread and/or influence of hate speech in real time. In a collaboration with Facebook, Bartlett and Krasodomski-Jones \shortcite{bartlett2015counter} observed counter-speech within political discussions on public pages across the network. They found that counter-speech operates differently according to the context in which it is produced and shared and also note significant methodological challenges inherent to determining when counter-speech is `successful’. Another study concluded that counter-speech is a more effective way of undermining hate speech than removing it \cite{benesch2016counterspeech}.

However, there is so far no consensus on how counter-speech can be detected online (e.g., through automated vs manual measures) \cite{faris2016understanding} and how its impact can be identified and measured. Focusing on Twitter, Wright et al. \shortcite{wright2017vectors} identified successful counter-speech through the existence of ‘golden conversations’ in which initial hate speech was countered and the poster(s) of that hate speech ultimately produced a response indicating some kind of favourable impact. Types of favourable impact included apologies, recanting of opinion, deletion of tweets and deletion of the user’s account. The authors acknowledge that the final two actions can be ambiguous in terms of meaning but nevertheless find that counter-speech can be effective in a range of social media environments, including in conversations between single and multiple users. Schieb and Preuss \shortcite{schieb2016governing} used a computational simulation model and gauged the effectiveness of counter-speech through measures such as the direction of influence of content achieved through user posts and liking of others’ posts. Applying this different methodological approach they also found indications that counter-speech can be effective. In particular, their findings suggested that even a small group of users can influence a larger audience if a significant number in that audience do not already possess extreme opinions.

In this paper we present the results of a micro-analytical study of a dataset of
interactional Twitter threads \cite{tolmie2018microblog}, that is a series of posts that are linked together by the use of the Twitter {\it reply} function. Each thread consisted of an initial or {\it source} post, followed by replies from other posters, either to the source post or to one another. We took thread length as a proxy for its potential for harmful impact and so our aim was to inform understanding of self-governance and cyber hate by investigating whether cyber hate posts attract support or disagreement by Twitter users and whether this has an impact on the length of those interactions. 


First, 
we have developed a statistical model of cyber hate Tweet interactions. 
Using thread length as its dependent measure and several independent variables that capture the nature of the subsequent responses within an individual cyber hate thread, we use this model to investigate factors that influence the volume of responses to cyber hate posts on Twitter. 
Second, 
we have trained a machine classifier to detect counter-speech. Using cyber hate and counter-speech classification in combination provides a way of automatically detecting and tracking cyber hate posts and counter-speech. 

\section{Examples of Cyber Hate and Counter-Speech}\label{cyberhateeg}
Understanding how social media users make cyber hate posts and counter-speech recognisable to others was important for our study in a number of ways. First, it provided material for training people to annotate our datasets. Second, it helped to identify content features that could be helpful in training a classifier to recognise counter-speech.

In the sample tweets below, we present brief extracts of Twitter interactional threads of cyber hate posts and counter-speech. We use a Conversation Analysis \cite{sacks1974simplest} approach to elucidate in detail what kinds of content might be considered by other users as antagonistic or hateful. The responses made to these posts also reveal what kinds of content could be seen as counter-speech.

We have chosen as examples two posts by Katie Hopkins, who tweets under the username @KTHopkins, together with two examples of the counter-speech they attracted.\footnote{There is debate over whether it is ethically appropriate to reproduce tweets in publications and bring them to the attention of a wider audience. As Katie Hopkins posts tweets as part of her work in the public eye, we do not consider it necessary to seek her consent to include them in further dissemination. The users who posted Tweets 3 and 4 were contacted over Twitter and gave consent for their posts to be included in our project publications. For ethics of publishing social media posts, see \cite{webb2017ethical}.} 
Hopkins is a newspaper columnist who is well known for expressing controversial opinions in an inflammatory style. 

\begin{figure}[ht]
\centering
\includegraphics[width=0.9\columnwidth]{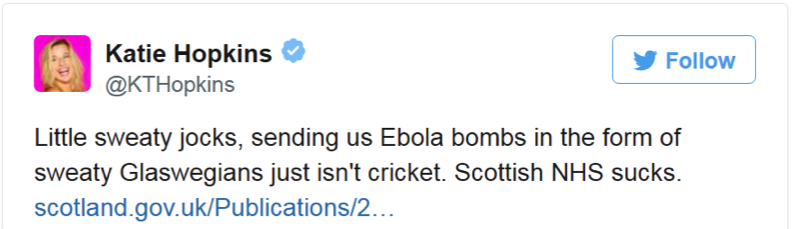}
\caption{Provocative tweet example 1.}
\label{fig:figure2}
\end{figure}

\begin{description} 
    \item[Tweet 1] The first example (Figure \ref{fig:figure2}) was posted in December 2014. 
    The tweet refers to news that a Scottish nurse who had contracted the Ebola virus after volunteering to work in Sierra Leone was being moved to an English hospital for specialist treatment. It makes a complaint about Scottish people receiving English care and does so in a way that can be seen to refer to the Scottish as inferior to the English. It begins with a reference to ``little sweaty jocks''. ``Jocks'' is a common colloquial term for Scottish people; it is not necessarily derogatory in meaning but is made so here through the descriptors ``little sweaty'' (with even a potential reference to a further meaning of `jocks' as `jockstrap').
    
    The tweet continues ``sending us Ebola bombs in the form of sweaty Glaswegians''. This refers to the movement of the nurse (who was from Glasgow) into an English hospital. The ``sweaty Glaswegians'' makes another derogatory reference -- to the nurse specifically and to Scottish people in general -- and is placed in opposition to ``us'' marking out a distinction between the Scottish and the English. ``Sending us Ebola bombs'' positions the Scottish as responsible for the spread of Ebola with the use of the verb ``sending'' suggesting that this is being done repeatedly and actively, even deliberately. ``Bombs'' marks this action as dangerous and destructive. The sentence is completed with a common English idiom referring to fairness, ``just isn't cricket'', and the text of the tweet concludes with a complaint that the ``Scottish NHS sucks''. 
    We note that the tweet does not include any individual words that can be seen as immediately offensive. Instead a derogatory category is constructed -- ``little sweaty jocks''-- and referred to negatively -- ``sending us Ebola bombs''. 
    
    
    \item[Tweet 2] The second example (Figure \ref{fig:figure3}) is constructed in a similar way. The tweet refers to ``ginger babies'' i.e. babies with ginger coloured hair. Once again, ``ginger'' is not necessarily an offensive term but is constructed as negative by further parts of the post. The ``like a baby'' marks them out as somehow different from other babies and carries a sense that this difference is for negative reasons. This sense is confirmed in the final part of the post which states that ginger babies are ``so much harder to love'' than other ones. As seen previously the tweet presents a personal opinion as fact and is positioned within a highly emotive context -- in this case the affection that most people feel for children. 

\begin{figure}[ht]
\centering
\includegraphics[width=0.9\columnwidth]{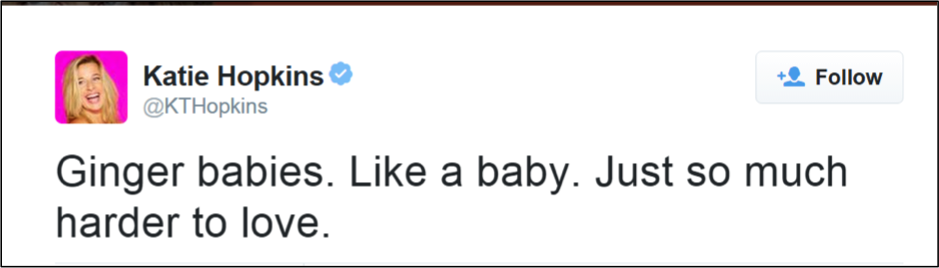}
\caption{Provocative tweet example 2.}
\label{fig:figure3}
\end{figure}

This brief examination demonstrates that tweets that might be considered antagonistic or hateful do not necessarily contain individual words that are offensive. Instead, inflammatory opinions are expressed by constructing negative references and drawing on emotive and/or divisive contexts. These observations help us towards a more nuanced understanding of the expression of cyber hate. Similar examination can highlight the specific ways that user responses to cyber hate posts are constructed and could be seen as performing the action of ‘counter-speech’. 

Both the Katie Hopkins tweets shown here received over 100 replies from other users. These expressed a range of responses including both support and counter-speech. Some treated the original posts as humorous whilst others made more serious responses that oriented to them as hateful and potentially damaging in impact. Tweets three (Figure \ref{fig:figure4}) and four (Figure \ref{fig:figure5}) are replies to the `ginger babies' tweet. Each post disagrees with and challenges the Katie Hopkins tweet and can be seen as a form of counter-speech. The placement of the user handle @KTHopkins at the start of the tweet indicates that the tweets are a reply and both users address the content of their tweet directly to Katie Hopkins through the use of `you'. 

\begin{figure}[h]
\centering
\includegraphics[width=0.9\columnwidth]{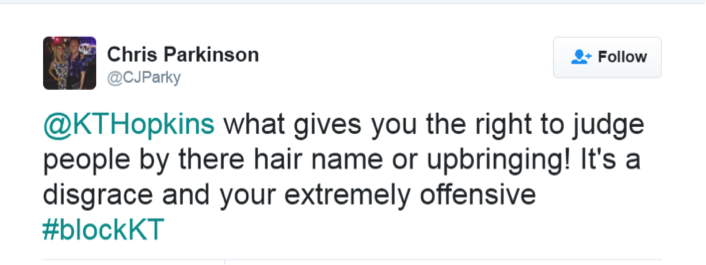}
\caption{counter-speech tweet example 1.}
\label{fig:figure4}
\end{figure}

\item[Tweet 3] The first example of a counter-speech tweet (Figure \ref{fig:figure4}) makes a reference to the content of the post it is replying to: ``What gives you the right to judge people by there [sic] hair name or upbringing!'' refers both to the subject matter of the tweet about `ginger babies' 
and its judgmental status. The formulation: ``What gives you the right \ldots'' is often used in counter-speech as a disagreement and challenge rather than a genuine question and this sense is confirmed here by the use of an exclamation mark (rather than question mark) at the end of the sentence and the subsequent statement: ``It's a disgrace and your [sic] extremely offensive''. 

\begin{figure}[h]
\centering
\includegraphics[width=0.9\columnwidth]{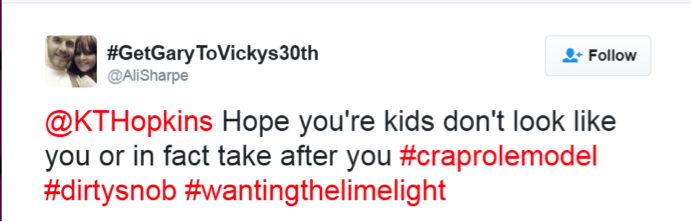}
\caption{counter-speech tweet example 2.}
\label{fig:figure5}
\end{figure}

\item[Tweet 4] The second example (Figure \ref{fig:figure5}) takes up the emotive context of `babies' by referring to Katie Hopkins' own children. The comment: ``Hope your kids don't look like or in fact take after you'' suggests that Katie Hopkins is unpleasant in appearance and personality -- as it would be unfortunate for her children to resemble her in either sense. We can note that this comment does not refer to the content on the original post directly but instead makes the user's disagreement with it clear via the critical evaluation of Katie Hopkins herself. Directly disagreeing responses (tweet 3) and indirect disagreements (tweet 4) that focus on the personality or appearance of the original posters are both frequently occurring replies in these kinds of threads and ones which we might regard as counter-speech. 

Another common feature of counter-speech replies is use of a hashtag to provide a further comment on the original post. These can often be seen to be directed to the wider readership of Twitter rather than (just) to the original poster and are therefore further actions of counter-speech. In tweet 3, `\#blockKT' can be seen as a kind of appeal or instruction to others to ignore Hopkins' posts and, in tweet 4, three separate hashstags appear to offer a commentary on Hopkins' status, values and motivation ``\#craprolemodel \#dirtysnob \#wantingthelimelight''. 
\end{description}

\section{Methodology}

\subsection{Data Collection}
The first step was to identify and collect examples of cyber hate threads on Twitter. To do this, we collected tweets from three Twitter `sentinel' accounts -- @Homophobes, @YesYoureRacist and @YesYoureSexist. Sentinel accounts set out to expose users posting what they judge to be cyber hate by encouraging counter-speech. We chose this approach because these accounts actively seek out and retweet hateful and antagonistic remarks that focus on people's personal and protected characteristics in order to highlight such posts to their followers and provoke counter-speech. Hence, these accounts are a suitable source of data for the purposes of examining and modeling responses to cyber hate posts. The first 100 tweets from each account were collected using a semi-automated process. A researcher clicked on each of the first 100 tweets, which linked to a new page within which the offending post and all subsequent replies were displayed. A PHP script was created to scrape this page and produce CSV files containing the text of the tweet and posting account. This produced a total of 300 files (100 homophobic, 100 racist, 100 sexist), each file containing an individual thread. Using examples of cyber hate posts and counter-speech (see subsection \ref{cyberhateeg}), the researcher collecting the data was briefed by experienced hate speech researchers on the indicative features of hateful remark to ensure a suitable corpus.

\subsection{Thread Annotation}
\begin{figure}[!ht]
\centering
\includegraphics[width=0.9\columnwidth]{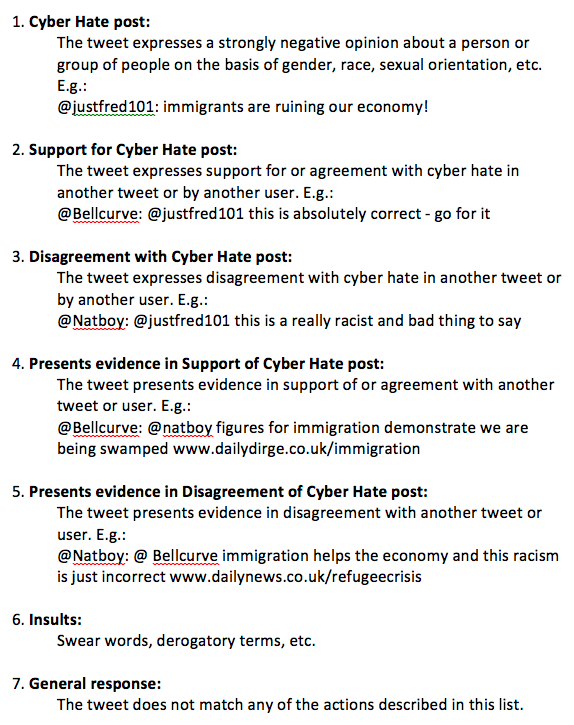}
\caption{Annotation scheme for cyber hate threads. More than one annotation code (1-6) can be applied to the training data.}
\label{fig:figure1}
\end{figure}

In designing the annotation scheme for posts in cyber hate threads, we drew on previous research to arrive at an appropriate methodology for the sociological analysis of micro-blogs, especially Twitter \cite{tolmie2018microblog}. This research identified a number of core organisational properties within Twitter thread interactions that have formed the basis of the approach reported here for annotating cyber hate threads, i.e., that interactions between users on Twitter have a sequential order; that they involve topic management; and that the production of a response is shaped by its accountable characteristics. The annotation scheme was designed via qualitative analysis of a sample of threads, drawing from Conversation Analysis \cite{sacks1974simplest}, Membership Categorisation Analysis \cite{hester1997culture,housley2017membership} and related principles that focused on speech acts \cite{housley2017digitizing}. The annotation scheme resulting from this iterative process is shown in Figure \ref{fig:figure1}. It assumes that there are two types of tweets in a thread: (1) the source tweet (cyber hate post) that initiates it and (2) response tweets that are replies within the thread started by the source tweet.\footnote{Tweets shown in Figure 5 are simulated examples.} 

We recruited an annotation team of four individuals to independently annotate each individual thread. The annotators were independent of the project to avoid selection bias and were well briefed on the instructions and process for annotation. This process involved opening each thread in a spreadsheet and, starting with the source tweet, annotating each replying tweet in order by inserting a number in the column next to each reply to indicate which code from the annotation scheme was the best match. Note the option was available for them to select `General response' if none of the other options matched. Annotators were also able to select more than one label per post. On completion of the annotation task we developed scripts to collate the annotations for the four independent annotators and determine levels of agreement between them.  

The aim of the quantitative analyses performed on the threads was (i) to determine the effect of different types of response to cyber hate with respect to thread length through statistical modeling, and (ii) to automate the process of response type identification by training a machine classifier to distinguish different responses and assign the correct label to each post. To establish confidence in the accuracy of annotations input to the statistical model and to the machine learning classifiers, we argue that it is the agreement score for the unit of analysis (each tweet), and not the overall annotator agreement for all units that is important for validation. Thus, we did not use general agreement scoring methods such as Krippendorf's Alpha or Cohen's Kappa; instead, we removed all tweets with less than 75 percent agreement and also those upon which the annotators could reach an absolute decision (i.e., the ‘undecided’ class). This approach has been used in related research \cite{ASI:ASI21180}. 


\subsection{Thread Length Modeling}
For thread length modelling we used a linear regression, where the dependent variable was thread length (number of posts in a thread including the original post). The independent variables used were counts of: hateful posts, supporting posts, disagreeing posts, insults, unique contributors, original poster contributions, unique contributors of hateful posts and unique contributors of counter-speech posts.

\subsection{Machine Classification of Responses}
The application of machine learning for classifying data requires identifying a set of features 
whose values are thought to be significant for determining the dependent variable (in this case, the response type). The first set of features we identified was a Bag of Words (BOW) \cite{wallach2006topic}. For each tweet the words were stemmed using the Snowball method and transformed to lowercase before being split into n grams of size 1-5, retaining 2,000 features, with word frequency normalised for each vector. 

The second set of features were extracted by identifying known hateful terms and phrases for cyber hate based on race, gender and sexual orientation. These were extracted from a crowd-sourced list of terms on Wikipedia.\footnote{https://en.wikipedia.org/wiki/Category:Slang\_terms\_for\_women}

The final set of features we identified were Typed Dependencies, which provide a representation of syntactic grammatical relationships in a sentence (or tweet) that can be used as features for classification and have the potential to capture othering language. Burnap et al. \shortcite{burnap2016us} identified that `othering' language (i.e., where some person or group is referred to in a way that implies they are ``not one of us'') was a useful feature for classifying cyber hate based on religion, race and sexuality. 
Based on our earlier analysis of the sample of Katie Hopkins' threads, we anticipated that this may also be present in responses to cyber hate. To extract potential othering terms we followed the same approach. Each tweet was transformed into typed dependency representation using the Stanford Lexical Parser \cite{stanford_dependencies}, along with a context-free lexical parsing model, transformed to lowercase, and split into n grams of size 1-3, retaining 2,000 features, with frequency normalisation applied.

Machine classification experiments were performed using (i) a Support Vector Machine (SVM) algorithm with a linear kernel and (ii) a Random Forest Decision Tree algorithm with 100 trees. The rationale for the selection of these machine learning algorithms was based on previous research that had analysed the performance of a range of alternative methods using similar data to those used in this study, and reported that these methods produced optimum results \cite{burnap2016us}. It was evident for the experiments performed in the present research that SVM continually outperformed the Random Forest approach, so only the SVM results are reported for brevity. Experiments were also conducted using RBF and Polynomial kernels, using SVM to establish whether a non-linear model fitted the data better, but both of these produced models with very poor detection of cyber hate. The SVM parameters were set to normalize data and use a gamma of 0.1 and C of 1.0, refined through experimentation.

\section{Results}
This section first summarises results of thread modeling and then of the response classification experiments.

\subsection{Thread Length Modeling}
Table \ref{tab:table1} shows the results of a linear regression using the count data of the independent variables and thread length as the dependent variable. There are a number of observations we can make that provide insight into the relationship between contributors' interactions within Twitter threads responding to hateful or antagonistic remarks, or cyber hate. The columns provide effect sizes (Coef) and statistical significance measures for features of cyber hate posts based on sexist, racist and homophobic bias. We refer to the three forms of bias collectively as strands. The total number of hateful remarks in a thread has no significant effect for any strand and support for the original hateful remark is only significant for racism. This suggests neither higher volume of cyber hate, nor increased support for cyber hate, influence the thread length. More contributors adding endorsement to cyber hate does not lead to a `mini wildfire'. However, counter-speech is statistically significant in explaining thread length for all strands, showing a positive association with thread length. The coefficients for all strands suggests more counter-speech leads to longer threads, which can be interpreted as attempted governance of such remarks by Twitter users increases threads. 

\begin{table*}[htb]
  \hfill{}
  \resizebox{\textwidth}{!}{%
  \begin{tabular}{|l|l|l|l|l|l|l|}
     \hline
\multicolumn{1}{|l|}{\multirow{2}{*}{}} & \multicolumn{2}{c|}{Sexist} &
\multicolumn{2}{c|}{Racist} & \multicolumn{2}{c|}{Homophobic} \\ \cline{2-7} 
\multicolumn{1}{|l|}{} & \multicolumn{1}{c|}{Coef.} & \multicolumn{1}{c|}{Std. Err.}  & \multicolumn{1}{c|}{Coef.} & \multicolumn{1}{c|}{Std. Err.} & \multicolumn{1}{c|}{Coef.} & \multicolumn{1}{c|}{Std. Err.}  \\ \hline
hatecount & -1.343604	& 3.543089 & -1.299386 & 1.247097 & 0.1803354 & 1.129952  \\  \hline
support & -1.647672	& 2.703798  & -3.458554**	 & 1.471944  & 0.0818545	& 0.6059934 \\  \hline
disagree & 4.638163***	& 1.105511 & 1.616641*& 0.9976502 & 1.436724*** & 0.3638562\\  \hline
insults & 2.371311	& 2.748198 &-2.105018	    & 5.257999& 1.127993***      &	0.3219493 \\  \hline
uniqcontributors & 2.874905***	& 0.3992179&1.686837***	    & 0.1315021 &	0.6387717***   &	0.0794272\\  \hline
origpostertweets & 1.545473***	& 0.2450241  &2.254713***	    & 0.1935309&	0.7773406***   &	0.1106594  \\  \hline
uniqhatefulcontributors & 0	(omitted) & & 0	(omitted)   & & -1.152764	    & 1.710793  \\  \hline
uniqCScontributors & -7.508296*** &	1.390719&  -2.313296*   &	1.316248 & -1.421774*** & 0.4042894 \\  \hline
cons & -3.883546	& 3.947113 &	     -2.842554***	& 1.05933 &	7.788954*** &	1.51231\\  \hline
Adj. R2 & 0.867	&  &	   0.6778	&  &	0.5521 &	\\  \hline
  \end{tabular}%
  }
  \caption{Thread Length Model - *p < 0.05; **p < 0.01; ***p < 0.001.}~
  \label{tab:table1}
\end{table*}

\begin{table*}[htb]
\hfill{}
\resizebox{\textwidth}{!}{%
\begin{tabular}{llllllllll}
\hline
\multicolumn{1}{|l|}{\multirow{2}{*}{}} & \multicolumn{3}{c|}{Sexist} &
\multicolumn{3}{c|}{Racist} & \multicolumn{3}{c|}{Homophobic} \\ \cline{2-10} 
\multicolumn{1}{|l|}{} & \multicolumn{1}{c|}{P} & \multicolumn{1}{c|}{R} & \multicolumn{1}{c|}{F} & \multicolumn{1}{c|}{P} & \multicolumn{1}{c|}{R} & \multicolumn{1}{c|}{F} & \multicolumn{1}{c|}{P} & \multicolumn{1}{c|}{R} & \multicolumn{1}{c|}{F} \\ \hline
\multicolumn{1}{|l|}{n-Gram words 1 to 5 with 2,000 features} & \multicolumn{1}{l|}{.759} & \multicolumn{1}{l|}{.782} & \multicolumn{1}{l|}{.770} & \multicolumn{1}{l|}{.736} & \multicolumn{1}{l|}{.771} & \multicolumn{1}{l|}{.753} & \multicolumn{1}{l|}{.794} & \multicolumn{1}{l|}{.849} & \multicolumn{1}{l|}{.820} \\ \hline
\multicolumn{1}{|l|}{n-Gram typed dependencies} & \multicolumn{1}{l|}{1.00} & \multicolumn{1}{l|}{.400} & \multicolumn{1}{l|}{.571} & \multicolumn{1}{l|}{.684} & \multicolumn{1}{l|}{.131} & \multicolumn{1}{l|}{.219} & \multicolumn{1}{l|}{.668} & \multicolumn{1}{l|}{.979} & \multicolumn{1}{l|}{.794} \\ \hline
\multicolumn{1}{|l|}{Combination of above} &  \multicolumn{1}{l|}{.754} & \multicolumn{1}{l|}{.782} & \multicolumn{1}{l|}{.768} &
\multicolumn{1}{l|}{.749} & \multicolumn{1}{l|}{.764} & \multicolumn{1}{l|}{.756} & \multicolumn{1}{l|}{.783} & \multicolumn{1}{l|}{.882} & \multicolumn{1}{l|}{.829} \\ \hline
\end{tabular}
}
\hfill{}
\caption{Precision, Recall and F-Measure results for counter-speech classifiers.}
\label{tab:table2}
\end{table*}

Interestingly, the results suggest that the number of unique users posting counter-speech is also statistically significant for all strands -- but is negatively associated with thread length for all strands. The reversal of effect in this case suggests that a smaller number of individuals interacting to express counter-speech is likely to extend the thread, but larger numbers of individuals performing counter-speech stems the thread. 
For completeness we note that the number of unique \textit{hateful} contributors was automatically omitted from the model for sexist and racist posts. As the number of hateful contributors was not statistically significant for homophobic posts we can assume its omission was due to the larger, significant effect sizes of the other variables. However, as already noted, the number of unique hateful contributors -- irrespective of the content of the responses -- was statistically significant across all strands and positively correlated with thread length. We would expect to see thread length increase as more users become involved, but this finding provides further weighting to the finding that an increase in unique counter-speech contributors has a curtailing effect on thread length. That is, more people becoming involved in responding to cyber hate increases the likelihood of a `mini wildfire', but mass self- governance through counter-speech reduces it.

\subsection{Machine Classification of Response Types}
Given the findings of the previous experiment on thread length modeling, 
we aimed to develop a method to automatically identify disagreement in responses to cyber hate. 
As with thread length, we developed methods to test the generality of the classifier across three types of bias. We produced results for (i) a baseline Bag of Words approach, (ii) probabilistic co-occurrences of ngrams using Typed Dependencies, and (iii) a combination of both. The results are shown in Table \ref{tab:table2}.

For this set of results a 10-fold cross-validation approach was used to train and test the supervised machine learning method. 
The results of the classification experiments are provided using standard text classification measures of: precision 
; recall 
and F-Measure.

\begin{table}[h]
\centering
\resizebox{\columnwidth}{!}{
\begin{tabular}{|l|l|l|l|l|}
\hline
 & Cyber & Support for & Disagree w/ & General\\
 & Hate & Hate & \& Insults & Response\\ \hline
 Sexist &  228 &  58  & 8 & 165 \\ \hline
 H'phobic  & 207 & 103 &  20 & 634  \\ \hline
 Racist  &  907 & 62 & 20 & 398  \\ \hline
\end{tabular}
}
\caption{Distribution of Annotations}
\centering
\label{tab:annotations}
\end{table}

\begin{table}[h]
\centering
\resizebox{\columnwidth}{!}{
\begin{tabular}{|l|l|l|l|l|}
\hline
 & Cyber & Support for & Disagree w/ & General\\
 & Hate & Hate & \& Insults & Response\\ \hline
 Cyber & & & &\\
 Hate & 206 & 1 & 0 & 21 \\ \hline
 Support & & & &\\
 for Hate & 4 & 35 & 0 & 19 \\ \hline
 Disagree w/ & & & & \\ 
 Hate \& Insults & 2 & 0 &  4 & 2 \\ \hline
 General & & & & \\ 
 Response & 26 & 10 & 0 & 129 \\ \hline
\end{tabular}
}
\caption{Sexist Confusion Matrix.}
\centering
\label{tab:table3}
\end{table}

Thread annotations were conflated during the classification stage due to very small numbers (less than 5 for all strands) of classes 4 and 5 of the annotation scheme -- evidence provided in support and disagreement. This left us with: cyber hate (class 0), support for cyber hate (class 1), disagreement with cyber hate and insults (class 2), and none of the above (class 3). Table \ref{tab:annotations} shows the distribution of annotations across all types of cyber hate. It is notable that disagreement and insults are relatively small within our sample. This makes the identification of distinguishing features of this type of language very challenging but ultimately we have a dataset that was robustly collected and resource intensive to develop, and we were testing the effectiveness of an annotation scheme developed through qualitative analysis of tweets. The small \textit{n} in this class suggests either the qualitative features used to produce the annotation scheme did not generalize across all tweets or that there is actually a lack of disagreement in response to cyber hate. Despite the  small \textit{n}, the overall performance of the classifier on all classes the results are encouraging, with F-measures of .768, .756 and .794 for sexist, racist and homophobic tweets respectively. The inclusion of Typed dependencies improves classification over the baseline Bag of Words approach for two of the tree classes, but the size of the improvement is small. If we look at the results on a class-by-class basis we can unpick a possible reason for the improvement being small and also study the value of this method for classifying counter-speech. Tables \ref{tab:table3}, \ref{tab:table4} and \ref{tab:table5} present the confusion matrices for each strand to illustrate classification error between the different classes. 

\begin{table}[h]
\centering
\resizebox{\columnwidth}{!}{
\begin{tabular}{|l|l|l|l|l|}
\hline
 & Cyber & Support for & Disagree w/ & General\\
 & Hate & Hate & \& Insults & Response\\ \hline
 Cyber & & & & \\ 
 Hate & 119 & 3 & 0 & 85 \\ \hline
 Support & & & & \\ 
 for Hate & 4 & 40  & 0 & 59 \\ \hline
 Disagree w/ & & & & \\
 Hate \& Insults & 6 & 3 & 0 & 11 \\ \hline
 General & & & & \\
 Response & 50 & 24 & 1 & 559\\ \hline
\end{tabular}
}
\caption{Homophobic Confusion Matrix.}
\label{tab:table4}
\end{table}

The annotated class label is in the left hand column, and the label the machine selected is in the top row. 
Overall, the results suggest the classifier performed well for class 0 (cyber hate), which we would expect given the findings of Burnap and Williams \cite{burnap2016us} who rigorously tested the Typed Dependencies approach. However, for classes 1 and 2 —- support for cyber hate posts and counter-speech  —- we see poor performance with up to 100\% error for counter-speech posts (class 2) in the homophobic strand (see Table \ref{tab:table4}). 

\begin{table}[h]
\centering
\resizebox{\columnwidth}{!}{
\begin{tabular}{|l|l|l|l|l|}
\hline
 & Cyber & Support for & Disagree w/ & General\\
 & Hate & Hate & \& Insults & Response\\ \hline
 Cyber & & & & \\
 Hate & 444 & 4 & 0 & 63\\ \hline
 Support & & & & \\ 
 for Hate & 4 & 30 & 1 & 27 \\ \hline
 Disagree w/ & & & & \\ 
 Hate \& insults & 4 & 2 & 2 & 12 \\ \hline
 General & & & & \\ 
 Response & 82 & 12 & 0 & 304\\ \hline
\end{tabular}
}
\caption{Racist Confusion Matrix.}
\label{tab:table5}
\end{table}

The support (class 1) classifier output performs significantly better, but still with error rates up to almost 70\% for the homophobic strand (only 40 out of 133 supportive tweets correctly classified). A possible explanation is that the value of Typed Dependencies comes from the identification of probabilistic occurrences of related terms and phrases. Burnap and Williams found this useful to detect `othering' language (e.g., send them home, get them out, burn it, etc.). The ``us and them'' is perhaps less likely in language surrounding disagreement and support in response to cyber hate and it is thus less likely that generalizable probabilistic language features such as othering would occur -- reducing the effectiveness of words and Typed Dependencies alone as features.

\balance
\section{Discussion}
Our study has examined one form of self-governance -- counter-speech -- and it has revealed that its effectiveness is sensitive to the way in which it is organised. To investigate if counter-speech is actually effective as a strategy for self-governance, we have presented a statistical model to explore factors that may influence the volume of responses to cyber hate posts. In particular, we found that the number of hateful posts is not a statistically significant measure of thread length for any type of cyber hate within the study. Thread length is increased by disagreement, so self-governance efforts prolong the thread. However, we also found that cyber hate thread length is negatively correlated with the number of unique counter-speech contributors to the thread.

We argue that these findings have some potentially significant implications for social media self-governance strategies for the second form of self-governance we observed in our study. Sentinel accounts set out to expose users posting what they judge to be cyber hate by encouraging counter-speech. Our findings suggest that to maximise effectiveness, sentinel accounts need to increase follower counts and find ways to encourage their active participation in challenging cyber hate.

While the evidence that encouraging counter-speech can be an effective mechanism for curbing cyber hate, at the same time it is necessary to be alert to the risks of promoting `digilantism' -- the aggressive shaming online of perceived wrong-doers \cite{prins2011digital}. Inevitably, therefore, questions of proportionality problematise assumptions of self-governance as an appropriate and effective counter-measure: how can its wider use be encouraged without risking descending into digilantism? 

Our machine classifier experiment results show reasonably effective performance across three types of cyber hate across all classes (cyber hate, support for cyber hate, counter-speech), but when looking in detail at the support and counter-speech classes, there is need for significant improvement in performance before this could be deployed alongside the statistical model. The nGram and Typed Dependency approach we implemented was possibly flawed due to generalization issues (with no semantic markup) and also the lack of any clear probabilistic co-occurrences of terms in the supporting or counter-speech tweets, which led to the success of classifying cyber hate in previous work.


The application of machine classification to counter-speech is intended to enable the automated detection of support and counter-speech in response to cyber hate. Given that we have already determined the statistically significant effect of counter-speech on curtailing thread length and stemming the potential risk of `mini wildfire', then such classifiers would provide `real-time' input into the statistical model (in terms of volumes of cyber hate, support, and counter-speech) as cyber hate tweets are posted to Twitter. 
This could assist organisations responsible for monitoring cyber hate 
to make more effective use of limited human resources.


\section{Conclusions and Future Work}

The results from our statistical modelling of the length of cyber hate threads have some useful implications for strategies for self-governance on social media. In particular, they reveal that the number of counter-speech posts, and the number of unique contributors to a thread, increase thread length of responses to cyber hate for sexist, racist and homophobic posts. While this is unsurprising given the nature of our dependent measure, what is interesting is the statistically significant finding that the number of unique counter-speech contributors has a negative effect on thread length, suggesting that engagement in counter-speech by more unique individual posters reduces the thread length, and thus is more effective for curtailing cyber hate.

The results of our experiments in applying machine classifiers to responses to cyber hate, together with the earlier work of Burnap and Williams \cite{burnap2016us}, suggest that their use in detecting automatically possible instances of cyber hate and subsequent counter-speech has a a role to play in social media governance strategies. At the very least, when used in combination, such classifiers will enable the closer study of self-governance -- how the Twitter user community reacts to support or denounce those posting cyber hate. The same approach also has the potential to help news media organisations who have had to remove the comments sections of their articles from their websites owing to resourcing issues surrounding the moderation of comments. 

However, further work is clearly required to improve machine classifier performance. This should focus on identifying interactional and linguistic features of cyber hate and counter-speech, for example, investigating the content and themes of 1-to-1, 1-to-few, and 1-to-many interactions in order to better understand their nature. We may assume from our results that such interactions could be argumentative or provocative where fewer numbers are involved, but overwhelming for the offensive poster where larger numbers respond with counter-speech leading to the cessation of further posts. 

So far, we have only taken the first steps in understanding how a detailed, qualitative analysis of interactional and linguistic features of twitter threads (see, e.g. tweets 3 and 4 in Section \ref{cyberhateeg}) could help to inform feature selection and thus improve classifier performance and these need to be investigated as part of a more detailed and interdisciplinary, thematic analysis of threads \cite{tolmie2018microblog,housley2017digitizing,housley2017membership}.

\section{Acknowledgements}
We wish to thank Alex Voss of the University of St Andrews for his help with harvesting Twitter threads. We would also like to thank members of various organisations who generously made time to talk with us. Finally, we would  like to thank the ESRC, the funders of the 'Digital Wildfires' project under Grant Number ES/L013398/1.

\bibliographystyle{acl_natbib}
\bibliography{cyberhate}

\begin{thebibliography}{21}
\expandafter\ifx\csname natexlab\endcsname\relax\def\natexlab#1{#1}\fi

\bibitem[{Awan(2014)}]{awan2014islamophobia}
Imran Awan. 2014.
\newblock Islamophobia and twitter: A typology of online hate against muslims
  on social media.
\newblock \emph{Policy \& Internet}, 6(2):133--150.

\bibitem[{Bartlett and Krasodomski-Jones(2015)}]{bartlett2015counter}
Jamie Bartlett and Alex Krasodomski-Jones. 2015.
\newblock Counter-speech examining content that challenges extremism online.
\newblock \emph{DEMOS, October}.

\bibitem[{Benesch et~al.(2016)Benesch, Ruths, Dillon, Saleem, and
  Wright}]{benesch2016counterspeech}
Susan Benesch, Derek Ruths, Kelly Dillon, Haji Saleem, and Lucas Wright. 2016.
\newblock Counterspeech on twitter: A field study. dangerous speech project.

\bibitem[{Burnap and Williams(2016)}]{burnap2016us}
Pete Burnap and Matthew~L Williams. 2016.
\newblock Us and them: identifying cyber hate on twitter across multiple
  protected characteristics.
\newblock \emph{EPJ Data Science}, 5(1):1.

\bibitem[{Faris et~al.(2016)Faris, Ashar, Gasser, and
  Joo}]{faris2016understanding}
Robert Faris, Amar Ashar, Urs Gasser, and Daisy Joo. 2016.
\newblock Understanding harmful speech online.
\newblock \emph{Berkman Klein Center Research Publication}, (2016-21).

\bibitem[{Gagliardone et~al.(2015)Gagliardone, Gal, Alves, and
  Martinez}]{gagliardone2015countering}
Iginio Gagliardone, Danit Gal, Thiago Alves, and Gabriela Martinez. 2015.
\newblock \emph{Countering online hate speech}.
\newblock UNESCO Publishing.

\bibitem[{Hester and Eglin(1997)}]{hester1997culture}
Stephen Hester and Peter Eglin. 1997.
\newblock \emph{Culture in action: Studies in membership categorization
  analysis}.
\newblock Number~4 in Studies in Ethnomethodology and Conversation Analysis.
  University Press of America.

\bibitem[{Housley et~al.(2017{\natexlab{a}})Housley, Webb, Edwards, Procter,
  and Jirotka}]{housley2017digitizing}
William Housley, Helena Webb, Adam Edwards, Rob Procter, and Marina Jirotka.
  2017{\natexlab{a}}.
\newblock Digitizing {S}acks? approaching social media as data.
\newblock \emph{Qualitative Research}.

\bibitem[{Housley et~al.(2017{\natexlab{b}})Housley, Webb, Edwards, Procter,
  and Jirotka}]{housley2017membership}
William Housley, Helena Webb, Adam Edwards, Rob Procter, and Marina Jirotka.
  2017{\natexlab{b}}.
\newblock Membership categorisation and antagonistic twitter formulations.
\newblock \emph{Discourse \& Communication}.

\bibitem[{de~Marneffe et~al.(2006)de~Marneffe, MacCartney, and
  Manning}]{stanford_dependencies}
Marie-Catherine de~Marneffe, Bill MacCartney, and Christopher~D. Manning. 2006.
\newblock \href {http://nlp.stanford.edu/pubs/LREC06_dependencies.pdf}
  {Generating typed dependency parses from phrase structure trees}.
\newblock In \emph{LREC}.

\bibitem[{Prins(2011)}]{prins2011digital}
Corien Prins. 2011.
\newblock Digital tools: Risks and opportunities for victims: Explorations in
  e-victimology.
\newblock In \emph{The New Faces of Victimhood}, pages 215--230. Springer.

\bibitem[{Sacks et~al.(1974)Sacks, Schegloff, and
  Jefferson}]{sacks1974simplest}
Harvey Sacks, Emanuel~A Schegloff, and Gail Jefferson. 1974.
\newblock A simplest systematics for the organization of turn-taking for
  conversation.
\newblock \emph{language}, pages 696--735.

\bibitem[{Schieb and Preuss(2016)}]{schieb2016governing}
Carla Schieb and Mike Preuss. 2016.
\newblock Governing hate speech by means of counterspeech on facebook.
\newblock In \emph{66th ica annual conference, at fukuoka, japan}, pages 1--23.

\bibitem[{Thelwall et~al.(2010)Thelwall, Wilkinson, and Uppal}]{ASI:ASI21180}
Mike Thelwall, David Wilkinson, and Sukhvinder Uppal. 2010.
\newblock \href {https://doi.org/10.1002/asi.21180} {Data mining emotion in
  social network communication: Gender differences in myspace}.
\newblock \emph{Journal of the American Society for Information Science and
  Technology}, 61(1):190--199.

\bibitem[{Titley et~al.(2014)Titley, Keen, and F{\"o}ldi}]{titley2014starting}
Gavan Titley, Ellie Keen, and L{\'a}szl{\'o} F{\"o}ldi. 2014.
\newblock Starting points for combating hate speech online.
\newblock \emph{Council of Europe, October 2014}.

\bibitem[{Tolmie et~al.(2018)Tolmie, Procter, Rouncefield, Liakata, and
  Zubiaga}]{tolmie2018microblog}
Peter Tolmie, Rob Procter, Mark Rouncefield, Maria Liakata, and Arkaitz
  Zubiaga. 2018.
\newblock Microblog analysis as a program of work.
\newblock \emph{ACM Transactions on Social Computing}, 1(1):2.

\bibitem[{Wallach(2006)}]{wallach2006topic}
Hanna~M Wallach. 2006.
\newblock Topic modeling: beyond bag-of-words.
\newblock In \emph{Proceedings of the 23rd international conference on Machine
  learning}, pages 977--984. ACM.

\bibitem[{Webb et~al.(2016)Webb, Burnap, Procter, Rana, Stahl, Williams,
  Housley, Edwards, and Jirotka}]{webb2016digital}
Helena Webb, Pete Burnap, Rob Procter, Omer Rana, Bernd~Carsten Stahl, Matthew
  Williams, William Housley, Adam Edwards, and Marina Jirotka. 2016.
\newblock {D}igital {W}ildfires: Propagation, verification, regulation, and
  responsible innovation.
\newblock \emph{ACM Transactions on Information Systems (TOIS)}, 34(3):15.

\bibitem[{Webb et~al.(2017)Webb, Jirotka, Stahl, Housley, Procter, Edwards,
  Williams, Rana, and Burnap}]{webb2017ethical}
Helena Webb, Marina Jirotka, Bernd Stahl, William Housley, Rob Procter, Adam
  Edwards, Matt Williams, Omer Rana, and Pete Burnap. 2017.
\newblock The ethical challenges of publishing twitter data for research
  dissemination.
\newblock In \emph{ACM Web Science}. ACM Press.

\bibitem[{Webb et~al.(2015)Webb, Jirotka, Stahl, Housley, Edwards, Williams,
  Procter, Rana, and Burnap}]{webb2015digital}
Helena Webb, Marina Jirotka, Bernd~Carsten Stahl, William Housley, Adam
  Edwards, Matthew Williams, Rob Procter, Omer Rana, and Pete Burnap. 2015.
\newblock '{D}igital {W}ildfires': a challenge to the governance of social
  media?
\newblock In \emph{Proceedings of the ACM web science conference}, page~64.
  ACM.

\bibitem[{Wright et~al.(2017)Wright, Ruths, Dillon, Saleem, and
  Benesch}]{wright2017vectors}
Lucas Wright, Derek Ruths, Kelly~P Dillon, Haji~Mohammad Saleem, and Susan
  Benesch. 2017.
\newblock Vectors for counterspeech on twitter.
\newblock In \emph{Proceedings of the First Workshop on Abusive Language
  Online}, pages 57--62.

\end{thebibliography}
\end{document}